\documentclass[conference, 10pt]{IEEEtran}
\ifCLASSINFOpdf
\else
\fi

\usepackage{graphicx}
\usepackage{color}
\usepackage{placeins}
\usepackage{float}
\usepackage{hyperref}
\usepackage{tabularx,colortbl}
\usepackage{url}      
\usepackage{amsmath}                           
\usepackage{array}
\begin{document}
%
\title{Equivalent Circuit Model for Thick Split Ring Resonators and Thick Spiral Resonators}
\author{\IEEEauthorblockN{L. M. Pulido-Mancera, J. D. Baena}
\IEEEauthorblockA{ Physics Department\\
 Universidad Nacional de Colombia, Bogot\'a, Colombia\\
 lmpulidom@unal.edu.co
}}
\maketitle

\begin{abstract}
A simple theoretical model which provides circuit parameters and resonance frequency of metallic thick resonators is presented. Two different topologies were studied: the original Pendry's SRR and spiral resonators of two and three turns. Theoretical computations of resonant frequencies are in good agreement with values obtained with a commercial electromagnetic solver. The model could be helpful for designing thick frequency selective surfaces (FSS) based on this types of resonators, so called metasurfaces.
\end{abstract}
\IEEEpeerreviewmaketitle

\section{Introduction}
The Split Ring Resonator (SRR) initially proposed by Pendry in 1999 \cite{SRR-Pendry} has been widely analyzed because of its small electrical size and strong magnetic polarizability at resonance which makes it a proper unit cell of metamaterial. For instance, the SRR joint to a metallic wire was used as the unit cell of the first experimental demonstration of a left-handed metamaterial by David Smith and co-workers in 2000 \cite{Smith2000}. After, particles showing similar behavior, like spiral resonators of two turns (SR2) \cite{Spiral-Baena}, have been also considered in order to reduce the electrical size of the unit cell. Main properties of these type of resonators can be approached by relatively easy LC-circuit models where the circuit parameters may be fully obtained from geometrical dimensions and material properties. In the case of infinitely thin resonators, readers could find some designer formulas in \cite{LC-circuit-Baena} for several different topologies. However, this calculations were only done for extremely thin resonator but some times it is interesting to manufacture thick structures in order to make the Q-factor higher.

\begin{figure}[h]
\begin{center}
\includegraphics[width=1\columnwidth]{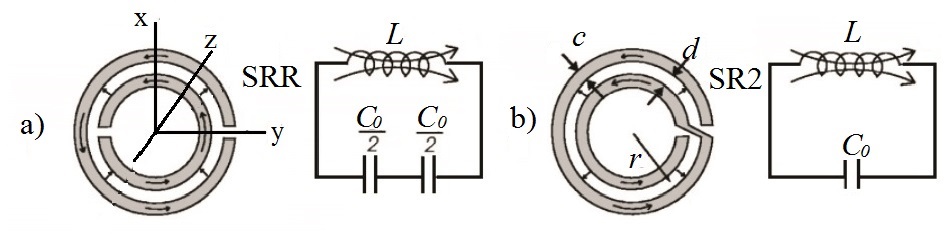} 
\caption{Resonant particles and its equivalent circuits: SRR (a) and SR2 (b). The arrows on the strips represent the current direction and the arrows within the gaps represent the displacement current. Both particles have the same average radius $r=$7.5mm and $c=d=$1mm.}
\label{fig1}
\end{center}
\end{figure}

In this paper a generalization of the LC-circuit model for thick resonators is presented. For instance, this model may be helpful for designing thick Frequency Selective Surfaces (FSS) with high quality factor based on these types of resonators, also known as metasurfaces.  Namely, along this paper, we focus our attention on thick particles with transverse shape of SRR and SR2 (see Fig.~\ref{fig1}).

\section{LC-circuit model for thick resonators}

\subsection{Calculation of L}

\begin{figure}
\centering
\includegraphics[width=0.8\columnwidth]{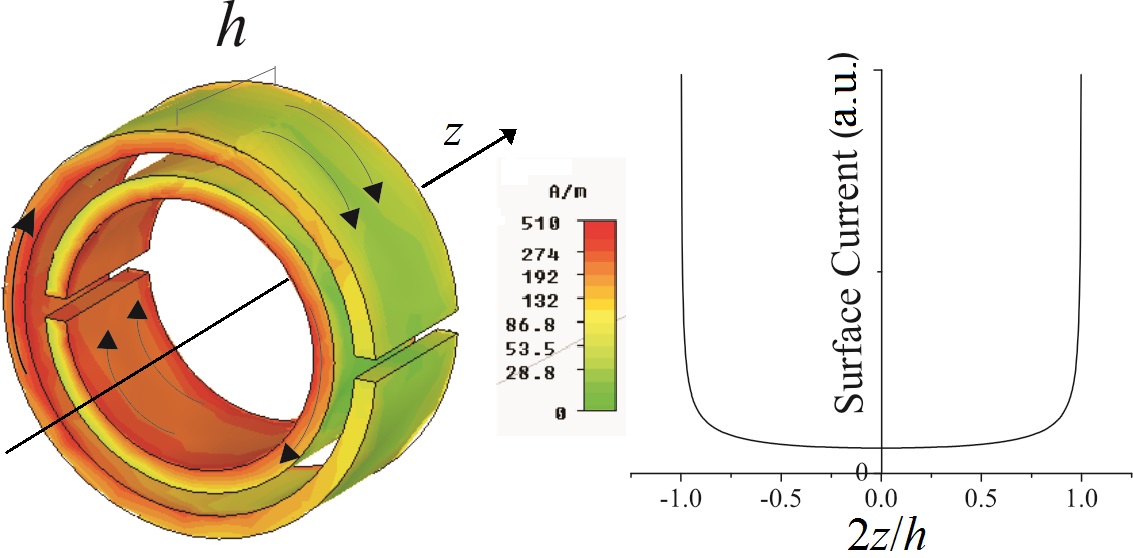} 
\caption{Maxwellian distribution of the current along the surface of the cylinder vs. thickness.}
\label{fig2}
\end{figure}

As the thickness $h$ is increased, although the induced current is spread along the $z$ direction, numerical simulations indicate that its profile on plane $xy$ approximately remains. Simulations also shows that near the borders of the metal strips the surface current diverges, so that it could be properly approximated by a Maxwellian distribution (see Fig.~\ref{fig2}). The inductance can be approached to that of a thin hollow metal cylinder of average radius $r$. The inductance of this ideal cylinder can be approximated by using the variational expression $L= 2U_{m}/I^{2}$, where $U_{m}$ is the magnetic energy. Assuming a Maxwellian distribution of the surface current density along the $z$ direction, we can get the following expression 
\begin{eqnarray}
\vec{J} (\rho, z) = J_{s\varphi}(z) \delta (\rho -r) \hat{\varphi} = \frac{J_{s0}}{\sqrt{1-(2z/h)^{2}}} \delta (\rho -r) \hat{\varphi}
\end{eqnarray} 
where $J_{s0}$ is the surface current density at $z=0$. The integration of (1) gives the current 
\begin{eqnarray}
I = \int_{-h/2} ^{h/2} \frac{J_{s0} dz}{\sqrt{1-(\frac{2z}{h})^{2}}} = J_{s0} \frac{h\pi}{2}
\end{eqnarray}
On the other hand, the magnetic energy is 
\begin{eqnarray}
U_{m} = \frac{1}{2} \int_{V} \vec{A}\cdot \vec{J} dV = \frac{1}{2} (2\pi r J_{s0}) \int_{-\infty}^{\infty} \frac{A_{\varphi}(r, z)dz}{\sqrt{1-(2z/h)^{2}}} \label{Um}
\end{eqnarray}
where the potential vector $\vec{A}$ can be calculated from the Laplace equation in cylindrical coordinates and corresponding boundary conditions:
\begin{eqnarray}
\frac{\partial^{2}A_{\varphi}}{\partial z^{2}} + \frac{\partial}{\partial \rho} (\frac{1}{\rho}\frac{\partial}{\partial \rho}(\rho A_{\varphi}))= 0 \label{laplace}\\
A_{\varphi}(r_{+},z)= A_{\varphi}(r_{-},z)\\ \label{bc-1}
\frac{\partial(\rho A_{\varphi})}{\partial \rho}|_{r_{+}} - 
\frac{\partial(\rho A_{\varphi})}{\partial \rho}|_{r_{-}} = r\mu_{0} J_{s\varphi}(r)
\end{eqnarray}
Using Fourier transforms for the surface current density and vector potential, equation (\ref{laplace}) turns into the Bessel equation 
\begin{eqnarray}
\frac{\partial^{2} \tilde{A_{\varphi}}}{\partial x^{2}} +  \frac{1}{x}\frac{\partial \tilde{A_{\varphi}}}{\partial x} - (1 + \frac{1}{x^{2}})\tilde{A_{\varphi}}=0, \label{bessel-equation}
\end{eqnarray}
where $x= k_{z}\rho$. The solution of this equation is 
\begin{eqnarray}
\tilde{A_{\varphi}}(r,k_{z})= \frac{J_{s0}\mu_{0} h}{i\sqrt{2\pi}}\frac{J_{0}(\frac{xh}{2r})I_{1}(x)K_{1}(x)}{k_{z}(K_{1}(x)I_{0}(x) + K_{0}(x)I_{1}(x))}
\end{eqnarray}
where $I_{n}(x)$ and $K_{n}(x)$ are the modified Bessel functions. Once the potential vector and the surface current are known, the variational expression $L= 2U_{m}/I^{2}$ can be used to get
\begin{eqnarray}
L = 2 r \mu_{0}\int_{0}^{\infty}\frac{(J_{0}(xh/2r))^{2}dx}{x(I_{0}(x)I_{1}^{-1}(x) + K_{0}(x)K_{1}^{-1}(x))} 
\end{eqnarray}

\subsection{Calculation of C}

In a first order approximation, the capacitance could be calculated by using the well known formula for the capacitance of a long cylindrical capacitor: 
\begin{eqnarray}
C_{h} = \frac{2 \pi h \epsilon}{\text{ln}(r/(r - d))};
\end{eqnarray}
where $h$ is the height or thickness, $d$ distance between the two coaxial cylinders, and $\epsilon$ the permittivity filling the space between the cylinders. However, for relatively small $h$ the fringing effect should not be neglected. We assume that the fringing electric field is similar to the field created by two coplanar extremely thin rings with the same shape of the transversal section of the cylindrical capacitor. The fringing capacitance may be approximated by $C_{0}= 2 \pi r C_{pul}$, where $C_{pul}$ represents the capacitance per unit of length of two parallel coplanar strips \cite{Bahl88}.

\subsection{The resonance frequency of SRR and SR2}

Considering that the two halves of the SRR are connected in series, the resonance frequency of its equivalent circuit is $\omega^{\text{SRR}}_{0}= (L(C_{h} + C_{0})/4)^{-1/2}$. On the other hand, the capacitance of the thick SR2 must be 4 times bigger, since the two halves are connected in parallel, i.e. $\omega^{\text{SR2}}_{0}= (L(C_{h} + C_{0}))^{-1/2}$. Therefore, the resonance frequency of SR2 is half the value for SRR, which means that the electrical size of the former is smaller than that of the last.

\section{Numerical validation}

\begin{figure}
\centering
\includegraphics[width=0.8\columnwidth]{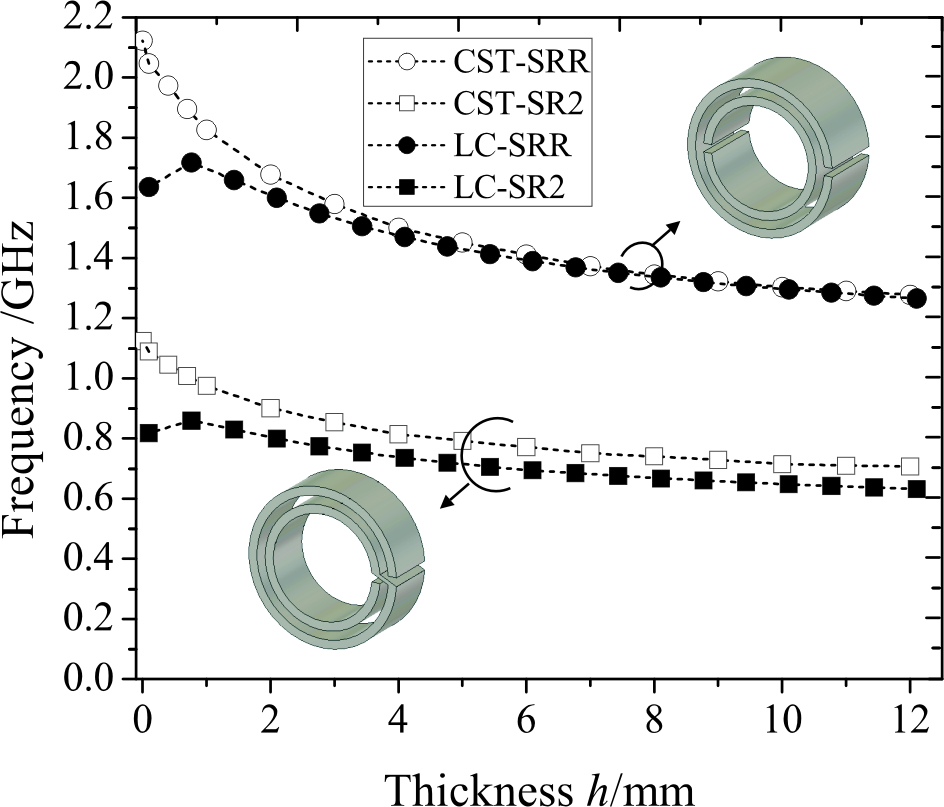} 
\caption{Resonance frequency vs. thickness $h$ calculated by means of the LC-circuit model and full-wave simulations by using \textit{CST Microwave Studio}.}
\label{fig3}
\end{figure}

In order to check the validity of the model, many numerical simulations were carried out by using the frequency domain solver of \emph{CST Microwave Studio}. All particles were made of perfect electric conductor in free space. The geometrical parameters on the transverse $xy$-plane were held constant and equal to the values indicated in caption of Fig.~\ref{fig1}. In Fig.~\ref{fig3} the resonance frequency has been plotted against the thickness $h$ varying within the range 0-12mm. This figure is also showing the values coming from the LC equivalent circuit model. For most part, the agreement between model is good. The disagreement observed for thin particles can be attributed to the fact that the approximation of L for a cylinder of infinitely thin metal is not longer valid, since $h$ start to be comparable to $c$ or $d$.

Besides, it is remarkable that the ratio of 2 between the resonance frequencies of thin SRR and SR2, which was previously reported \cite{Spiral-Baena}, is still obtained for thick particles.

\section{Conclusions}
A circuit model for calculating the resonance frequency of thick SRR and SR2 has been proposed and checked by means of numerical simulations. Besides, it was shown that the ratio between the resonance frequency of the SRR and the SR2 is independent of the thickness and approximately equal to 2, similarly to the case of thin particles \cite{Spiral-Baena}. This analytical model can be useful for designing thick FSS with higher Q-factor than that of thin particles.

\end{document}